\documentclass[pdflatex,sn-aps]{sn-jnl}

\usepackage{graphicx}
\usepackage{color}
\usepackage{amsthm,amssymb,amsmath}
\usepackage{amsfonts}
\usepackage{mathtools}
\usepackage{bm}
\usepackage{physics, ulem}
\usepackage{comment}
\usepackage{ascmac}
\usepackage{empheq}
\usepackage[thicklines]{cancel}
\usepackage{enumerate}
\usepackage{wrapfig}
\usepackage[version=4]{mhchem}
\usepackage{cases}
\usepackage{hyperref}

\usepackage{multirow}
\usepackage{mathrsfs}
\usepackage[title]{appendix}
\usepackage{xcolor}
\usepackage{textcomp}
\usepackage{manyfoot}
\usepackage{booktabs}
\usepackage{algorithm}
\usepackage{algorithmicx}
\usepackage{algpseudocode}
\usepackage{listings}
\usepackage{orcidlink}

\usepackage{caption}
\usepackage[subrefformat=parens]{subcaption}
\hypersetup{
citecolor = blue,
colorlinks = true,
urlcolor = blue
}

\theoremstyle{thmstyleone}

\theoremstyle{thmstyletwo}

\theoremstyle{thmstylethree}

\allowdisplaybreaks

\newcommand{\Def}{\coloneq}
\newcommand{\deF}{\eqcolon}
\newcommand{\up}{\uparrow}
\newcommand{\down}{\downarrow}

\newcommand{\R}{\mathbb{R}}

\newcommand{\Z}{\mathbb{Z}}
\newcommand{\kx}{k_\mathrm{x}}
\newcommand{\ky}{k_\mathrm{y}}
\newcommand{\kz}{k_\mathrm{z}}
\newcommand{\Pf}{\mathrm{Pf}}

\begin{document}

\title[Superconducting Gap Structures in Wallpaper Fermion Systems]{Superconducting Gap Structures in Wallpaper Fermion Systems}

\author*[1]{\fnm{Kaito} \sur{Yoda} \orcidlink{0009-0006-5607-6364}}\email{yoda.kaito.c1@s.mail.nagoya-u.ac.jp}

\author[1]{\fnm{Ai} \sur{Yamakage} \orcidlink{0000-0003-4052-774X}}\email{yamakage.ai.x6@f.mail.nagoya-u.ac.jp}
\equalcont{This author contributed equally to this work.}

\affil*[1]{\orgdiv{Department of Physics}, \orgname{Nagoya University}, \orgaddress{\street{Furo-cho, Chikusa-ku}, \city{Nagoya}, \postcode{4648602}, \state{Aichi}, \country{Japan}}}

\abstract{
We theoretically investigate the superconducting gap structures in wallpaper fermions, which are surface states of topological nonsymmorphic crystalline insulators, based on a two-dimensional effective model.
A symmetry analysis identifies six types of momentum-independent pair potentials.
One hosts a point node, two host line nodes, and the remaining three are fully gapped.
By classifying the Bogoliubov--de Gennes Hamiltonian in the zero-dimensional symmetry class, we show that the point and line nodes are protected by $\Z_2$ topological invariants.
In addition, for the twofold-rotation-odd pair potential, nodes appear on the glide-invariant line and are protected by crystalline symmetries, as clarified by the Mackey--Bradley theorem.
}

\keywords{Superconductivity, Topological Crystalline Insulator, Wallpaper Fermion, Superconducting Gap Structure}

\maketitle

\section{Introduction}\label{sec1}

Topological superconductivity in topological materials has been extensively studied in the past few decades, such as topological insulators~\cite{yonezawa2018nematic}, topological semimetals~\cite{kobayashi2015topological, lu2015crossed, aggarwal2016unconventional, wang2016observation, wang2017discovery}, and topological crystalline insulators~\cite{sasaki2012odd, novak2013unusual, he2013full, hashimoto2015surface, kawakami2018topological}.
In such systems, crystalline symmetries can forbid the opening of a superconducting gap in Dirac fermions on the surface, allowing the coexistence of the Dirac and Majorana fermions.
These two fermions hybridize with each other, and their energy spectrum exhibits features distinct from a simple linear dispersion.
Notably, at a specific parameter including the chemical potential, the system undergoes the Lifshitz transition \cite{lifshitzanomalies}, where the group velocity is zero, resulting in a divergence in the density of states~\cite{hao2011surface, hsieh2012Majorana, yamakage2012theory, yamakage2013theory}.
Near this critical point, surface-sensitive physical quantities, such as the tunneling conductance in normal metal/superconductor junctions~\cite{yamakage2012theory, yamakage2013theory} and Josephson currents~\cite{yamakage2013anomalous}, are significantly enhanced.
Such a feature is characteristic of topological superconductors whose parent normal state is a topological material and does not occur in topological superconductivity from topologically trivial metals.
Moreover, while similar mechanism-driven phenomena can be expected in many topological materials, studies to date have been conducted only on systems with symmorphic crystalline symmetry.

Our research focuses on superconductivity in topological nonsymmorphic crystalline insulators with a wallpaper fermion on the surface~\cite{wieder2018wallpaper, ryu2020wallpaper, zhou2021glide, hwang2023magnetic, mizuno2023hall, mizuno2025magnon}, which hosts fourfold degeneracy protected by time-reversal and two orthogonal glide symmetries.
The multiple energy spectra protected by the nonsymmorphic crystalline symmetry possess properties distinct from Dirac fermions on the surface of topological insulators~\cite{mizuno2023hall}.
Thus, the hybridization between wallpaper and Majorana fermions in topological superconducting states may give rise to novel quasiparticles unique to nonsymmorphic systems.

Observation of hybridization between surface wallpaper and Majorana fermions requires that the wallpaper fermions remain gapless in the superconducting state.
Here, we determine, for momentum-independent pair potentials characteristic of weakly correlated systems, the resulting gap structure of the wallpaper fermions.
In addition, we elucidate the underlying mechanisms of the nodal structures through the use of topological invariants and group-theoretical analysis.

The paper is structured as follows.
In Sec.~\ref{sec2}, we introduce an effective model for wallpaper fermions~\cite{mizuno2023hall} as the normal Hamiltonian and derive possible pair potentials based on crystalline symmetry.
Section~\ref{sec3} presents numerical results for superconducting gap structures in wallpaper fermion systems, and we elucidate the theoretical mechanisms underlying the emergence of the nodal structures, using the zero-dimensional topological invariants and the group-theoretical approach.
Section~\ref{sec4} discusses the distinction between superconducting gap nodes protected by topological invariants and those protected by crystalline symmetries.
Finally, Sec.~\ref{sec5} provides a summary of the main findings.

\section{Pair Potentials for Wallpaper Fermions}\label{sec2}

\subsection{Wallpaper Fermions}\label{subsec2-1}

Wallpaper fermions with fourfold degeneracy emerge on the surface of topological nonsymmorphic crystalline insulators that have two orthogonal glides, i.e., with symmetry of wallpaper groups $\mathrm{p4g}$ or $\mathrm{pgg}$.
We focus on the higher-symmetric one, $\mathrm{p4g}$, which has fourfold-degenerate states at the $\bar{\mathrm{M}}$ point and twofold-degenerate states on the $\bar{\mathrm{X}} \bar{\mathrm{M}}$ line~\cite{wieder2018wallpaper}.

\begin{figure}[t]
    \centering
    \includegraphics[keepaspectratio,scale=0.8]{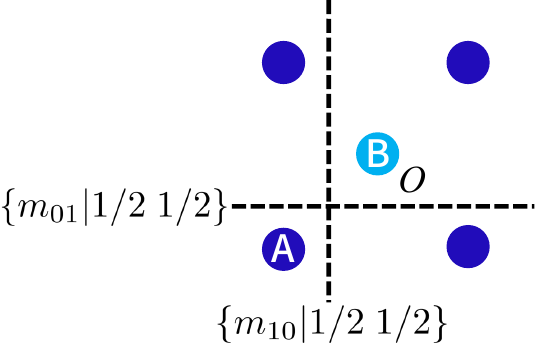}
    \caption{(Color online)
            Crystal structure protecting wallpaper fermions.
            This crystalline symmetry corresponds to the wallpaper group $\mathrm{p4g}$.
            Two glides $\{m_{01}|1/2\ 1/2\}$ and $\{m_{10}|1/2\ 1/2\}$ are indicated by the dashed lines.
            }
    \label{fig:p4g}
\end{figure}

\begin{figure}[t]
    \centering
    \begin{tabular}{cc}
        \begin{minipage}[t]{0.4\hsize}
            \centering
            \includegraphics[keepaspectratio,scale=0.4]{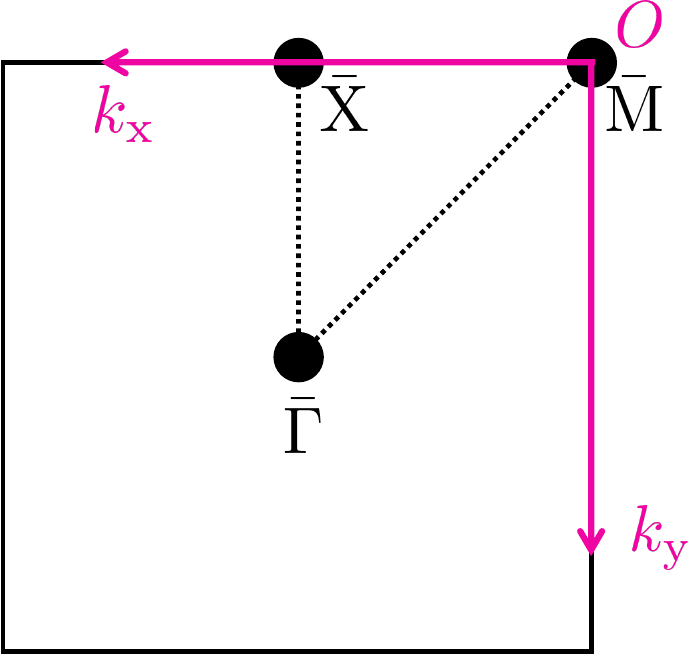}
            \caption{(Color online)
                    Brillouin zone of the wallpaper group $\mathrm{p4g}$.
                    Hereafter, we set the origin at $\bar{\mathrm{M}}$ point.
                    }
            \label{fig:BZ}
        \end{minipage} &
        \begin{minipage}[t]{0.53\hsize}
            \centering
            \includegraphics[keepaspectratio,scale=0.23]{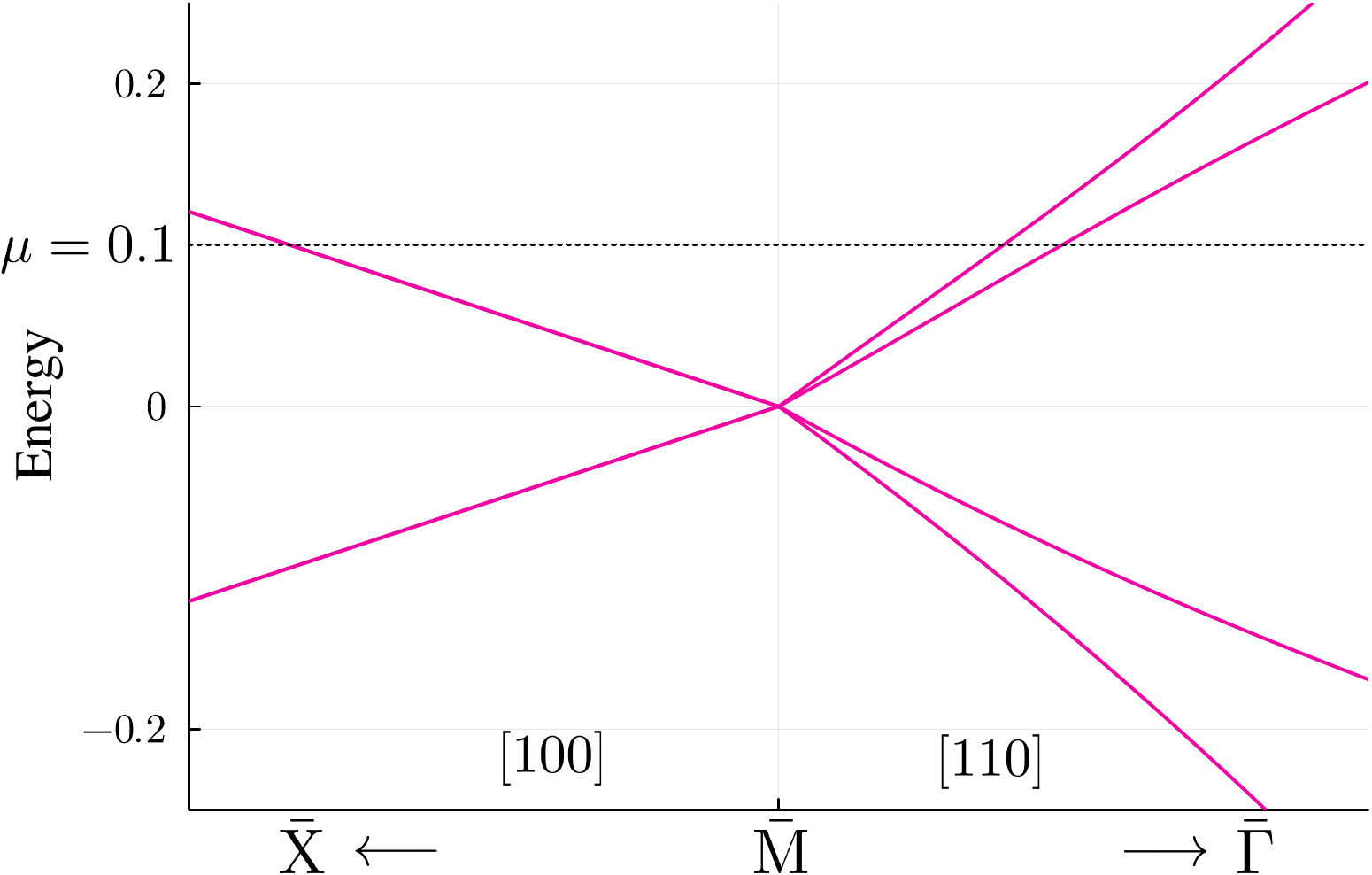}
            \caption{(Color online)
                    Energy dispersion of a wallpaper fermion with $v_1=-0.099,\ v_2=-0.208,\ v_3=0.066,\ v_4=0.127,\ v_5=0.251$.
                    }
            \label{fig:wp_energy}
        \end{minipage}
    \end{tabular}
\end{figure}

In the concrete analysis, we employ a simple model defined on the crystal structure~\cite{wieder2018wallpaper}, with the lattice constant set to unity, as shown in Fig.~\ref{fig:p4g}.
We begin by introducing the effective Hamiltonian for wallpaper fermions.
The symmetry operations $g$ and their representation matrices $D(g)$ in the little group at the $\bar{\mathrm{M}}$ point of the wallpaper group $\mathrm{p4g}$ are summarized in Table~\ref{tab:rep_mat}
\footnote{
        In Ref.~\cite{mizuno2023hall}, they derived the representation matrices of the space group $\mathrm{P4bm}$, which is equivalent to the wallpaper group $\mathrm{p4g}$ except for the translation along the $\mathrm{z}$ direction.
        }.
\begin{table*}
    \centering
    \caption{
            The symmetry operations $g$ and their representation matrices $D(g)$ in the little group at the $\bar{\mathrm{M}}$ point of the wallpaper group $\mathrm{p4g}$~\cite{Aroyo2011-cr, Aroyo2006-bi1, Aroyo2006-bi2, elcoro2021magnetic, xu2020high}.
            The phase factors $D(g)$ are fixed so as to commute with time reversal $\Theta$, $[D(g),\Theta]=0$.
            The representations of generators $D(\{4^{+}|\bm{0}\})$ and $D(\{m_{01}|1/2\ 1/2\})$ of the wallpaper group $\mathrm{p4g}$ are derived in Ref.~\cite{mizuno2023hall}.
            }
    \begin{tabular}{ll}
        \hline\hline
        $g$ & $D(g)$ \\ \hline
        $\{E|\bm{0}\}$ & $s_0\sigma_0$ \\
        $\{4^{+}|\bm{0}\}$ & $-\frac{1}{\sqrt{2}}(s_0\sigma_\mathrm{z}-is_\mathrm{z}\sigma_0)$ \\
        $\{4^{-}|\bm{0}\}$ & $-\frac{1}{\sqrt{2}}(s_0\sigma_\mathrm{z}+is_\mathrm{z}\sigma_0)$ \\
        $\{2|\bm{0}\}$ & $-is_\mathrm{z}\sigma_\mathrm{z}$\\
        $\{m_{10}|1/2\ 1/2\}$ & $-\frac{1}{\sqrt{2}}(s_0\sigma_\mathrm{x}-s_\mathrm{z}\sigma_\mathrm{y})$ \\
        $\{m_{01}|1/2\ 1/2\}$ & $-\frac{1}{\sqrt{2}}(s_0\sigma_\mathrm{x}+s_\mathrm{z}\sigma_\mathrm{y})$ \\
        $\{m_{11}|1/2\ 1/2\}$ & $-is_0\sigma_\mathrm{y}$ \\
        $\{m_{1\bar{1}}|1/2\ 1/2\}$ & $is_\mathrm{z}\sigma_\mathrm{x}$ \\ \hline\hline
    \end{tabular}
    \label{tab:rep_mat}
\end{table*}
Here, $s_\nu$ and $\sigma_\nu$ ($\nu=0,\mathrm{x},\mathrm{y},\mathrm{z}$) denote the Pauli matrices in the spin and sublattice spaces, respectively.
By imposing invariance under these symmetry operations, $D(g) H_{\mathrm{wp}}^{(\mathrm{eff})}(\bm{k}) D(g)^\dagger = H_{\mathrm{wp}}^{(\mathrm{eff})}(g\bm{k})$, together with the time-reversal $\Theta\Def-is_\mathrm{y}\sigma_0\mathcal{K}$, where $\mathcal{K}$ denotes complex conjugation, the effective Hamiltonian expanded up to $\bm{k}^2$ is obtained as follows:
\begin{equation}
    \begin{split}
        H^{(\mathrm{eff})}_{\mathrm{wp}}(\bm{k})\Def
        &\frac{1}{\sqrt{2}}\big[v_1\qty{(s_{\mathrm{z}}\sigma_{\mathrm{x}}+s_{0}\sigma_{\mathrm{y}})k_{\mathrm{x}} + (s_{\mathrm{z}}\sigma_{\mathrm{x}}-s_{0}\sigma_{\mathrm{y}})k_{\mathrm{y}}} \\
        &+v_2\qty{(s_{\mathrm{y}}\sigma_{\mathrm{z}}+s_{\mathrm{x}}\sigma_0)k_{\mathrm{x}} + (s_{\mathrm{y}}\sigma_{\mathrm{z}}-s_{\mathrm{x}}\sigma_0)k_{\mathrm{y}}} \\
        &+v_3\qty{(s_{\mathrm{x}}\sigma_{\mathrm{z}}-s_{\mathrm{y}}\sigma_0)k_{\mathrm{x}} + (s_{\mathrm{x}}\sigma_{\mathrm{z}}+s_{\mathrm{y}}\sigma_0)k_{\mathrm{y}}}\big] \\
        &+(v_4 s_\mathrm{x}\sigma_\mathrm{y}+v_5 s_\mathrm{y}\sigma_\mathrm{y})k_{\mathrm{x}}k_{\mathrm{y}},
    \end{split}
    \label{eq:wp}
\end{equation}
where $v_i\,(i = 1,\ldots,5)\in\R$.
The basis is chosen as $(c_{\mathrm{A},\up},c_{\mathrm{B},\up},c_{\mathrm{A},\down},c_{\mathrm{B},\down})$, and the origin and axes in the momentum space are defined as shown in Fig.~\ref{fig:BZ}.
The corresponding energy dispersion is plotted in Fig.~\ref{fig:wp_energy}.
This Hamiltonian~\eqref{eq:wp} is adopted as the normal-state model in the remainder of this work.

\subsection{Classification of the pair potential}\label{subsec2-2}
To describe the superconducting wallpaper fermions, we first define the Bogoliubov--de Gennes (BdG) Hamiltonian as
\begin{equation}
    H^{(\mathrm{eff})}_\mathrm{scwp}(\bm{k})\Def
    \qty[H^{(\mathrm{eff})}_\mathrm{wp}(\bm{k})-\mu]\tau_\mathrm{z}+\hat{\Delta}_i\tau_\mathrm{x},
    \label{eq:BdG}
\end{equation}
where $\tau_\nu \ (\nu=0,\mathrm{x},\mathrm{y},\mathrm{z})$ is the Pauli matrix in the particle-hole (Nambu) space, $\mu$ is the chemical potential, and $\hat{\Delta}_i$ denotes the pair potential.
The basis is taken as $(c_{X,\up},c_{X,\down},-c^\dagger_{X,\down},c^\dagger_{X,\up})$ with $X=\mathrm{A},\mathrm{B}$.

We identify the possible types of the pair potential $\hat{\Delta}_i$.
We assume that $\hat{\Delta}_i$ is momentum independent in this basis, and impose the Fermi--Dirac statistics $\hat{\Delta}_i=s_\mathrm{y}\hat{\Delta}_i^\mathrm{t}s_\mathrm{y}$.
As shown in Table~\ref{tab:Delta}, we can decompose the pair potentials into the irreducible representations (irreps) of the point group $\mathrm{C}_{4\mathrm{v}}$, which is the rotational part of the wallpaper group $\mathrm{p4g}$.
\begin{table*}[t]
    \centering
    \caption{
            Decomposition of pair potentials into the irreducible representations (irreps) of the point group $\mathrm{C}_{4\mathrm{v}}$.
            We assume momentum-independent pairings and impose Fermi statistics.
            ${}^1\mathrm{E}$ $({}^2\mathrm{E})$ represents the first (second) component of $\mathrm{E}$ representation corresponding to $\kx$ ($\ky$).
            $\Delta_0$ is a constant independent of momentum.
            }
    \begin{tabular}{crrrrrcl}
        \hline\hline
        Irreps & $E$ & $2C_{4\mathrm{z}}$ & $C_{2\mathrm{z}}$ & $2\sigma_\mathrm{v}$ & $2\sigma_\mathrm{d}$ & Basis & Pair potential \\ \hline
        $\mathrm{A}_1$& $1$ & $1$ & $1$ & $1$ & $1$ & $\kz$ &
        $\hat{\Delta}_1\Def \Delta_0s_0\sigma_0$ \\
        $\mathrm{A}_2$ & $1$ & $1$ & $1$ & $-1$ & $-1$ & $\kx\ky(\kx^2-\ky^2)$ &
        $\hat{\Delta}_2\Def \Delta_0s_0\sigma_{\mathrm{z}}$ \\
        $\mathrm{B}_1$ & $1$ & $-1$ & $1$ & $1$ & $-1$ & $ \kx^2-\ky^2 $ & \\
        $\mathrm{B}_2$ & $1$ & $-1$ & $1$ & $-1$ & $1$ & $ \kx\ky $ &
        $\hat{\Delta}_3\Def \Delta_0s_\mathrm{x}\sigma_\mathrm{y}$, \\
        & & & & & & & $\hat{\Delta}_4\Def \Delta_0s_\mathrm{y}\sigma_\mathrm{y}$ \\
        $\mqty{{}^1\mathrm{E} \\ {}^2\mathrm{E}}$ & $2$ & $0$ & $-2$ & $0$ & $0$ & $\mqty{\kx \\ \ky}$ &
        $\mqty{
                \hat{\Delta}_5\Def \Delta_0(s_0\sigma_\mathrm{x}+s_\mathrm{z}\sigma_\mathrm{y}) \\
                \hat{\Delta}_6\Def \Delta_0(s_0\sigma_\mathrm{x}-s_\mathrm{z}\sigma_\mathrm{y})
                }$ \\ \hline\hline
    \end{tabular}
    \label{tab:Delta}
\end{table*}
Thus, there are six possible momentum-independent pairings $\Delta_1$--$\Delta_6$ shown in Table~\ref{tab:Delta}.
For each of these pair potentials, we will calculate the corresponding nodal structure.

\section{Superconducting gap structures}\label{sec3}

\subsection{Numerical results}\label{subsec3-1}
We present the numerical results of superconducting gap structures in wallpaper fermion systems.
Throughout the calculations, we set the magnitude of the pair potential in Table~\ref{tab:Delta} to $\Delta_0 = 0.005$, assuming the weak-coupling pairing.
$v_i\,(i=1,\ldots,5)$ is set at the same value as shown in Fig.~\ref{fig:wp_energy} and the chemical potential is fixed at $\mu = 0.1$.

\begin{figure}[t]
    \centering
    \begin{tabular}{ccc}
        \begin{minipage}[t]{0.3\hsize}
            \centering
            \captionsetup{position=top, justification=raggedright, singlelinecheck=false}
            \subcaption{}
            \includegraphics[keepaspectratio, scale=0.6]{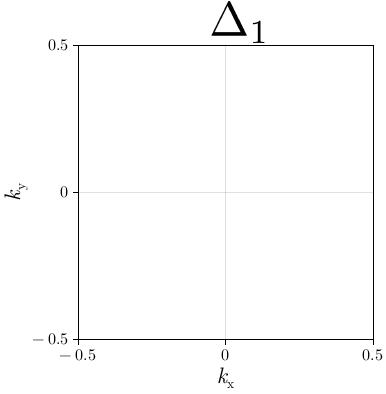}
            \label{fig:node_delta_1}
        \end{minipage} &

        \begin{minipage}[t]{0.3\hsize}
            \centering
            \captionsetup{position=top, justification=raggedright, singlelinecheck=false}
            \subcaption{}
            \includegraphics[keepaspectratio, scale=0.6]{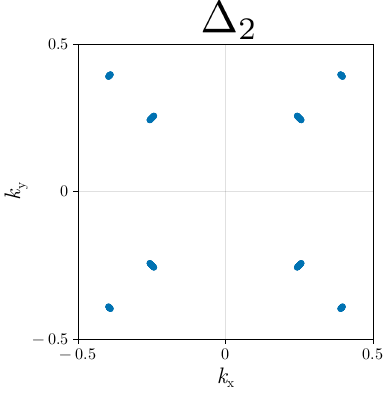}
            \label{fig:node_delta_2}
        \end{minipage} &

        \begin{minipage}[t]{0.3\hsize}
            \centering
            \captionsetup{position=top, justification=raggedright, singlelinecheck=false}
            \subcaption{}
            \includegraphics[keepaspectratio, scale=0.6]{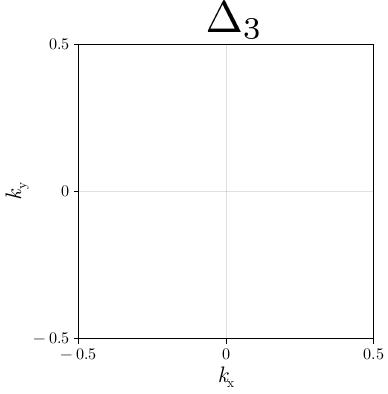}
            \label{fig:node_delta_3}
        \end{minipage} \\

        \begin{minipage}[t]{0.3\hsize}
            \centering
            \captionsetup{position=top, justification=raggedright, singlelinecheck=false}
            \subcaption{}
            \includegraphics[keepaspectratio, scale=0.6]{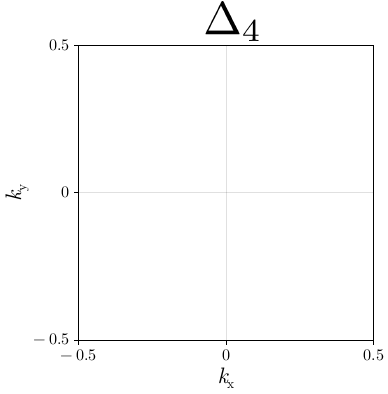}
            \label{fig:node_delta_4}
        \end{minipage} &

        \begin{minipage}[t]{0.3\hsize}
            \centering
            \captionsetup{position=top, justification=raggedright, singlelinecheck=false}
            \subcaption{}
            \includegraphics[keepaspectratio, scale=0.6]{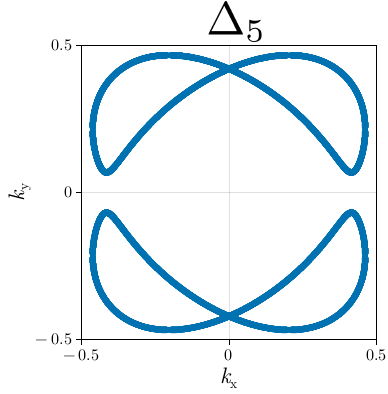}
            \label{fig:node_delta_5}
        \end{minipage} &

        \begin{minipage}[t]{0.3\hsize}
            \centering
            \captionsetup{position=top, justification=raggedright, singlelinecheck=false}
            \subcaption{}
            \includegraphics[keepaspectratio, scale=0.6]{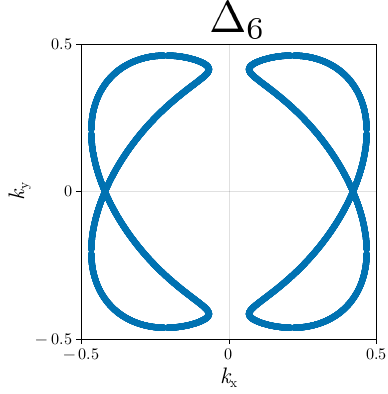}
            \label{fig:node_delta_6}
        \end{minipage} \\
    \end{tabular}
    \caption{(Color online)
            Nodal structures in superconducting wallpaper fermions.
            We identify the momentum points where the energy is smaller than $10^{-4}$ as gapless points.
            We set $\mu=0.1$ and assume the weak-coupling pairing ($\Delta_0=0.005$).
            The other parameters are the same values used in Fig.~\ref{fig:wp_energy}.
            }
    \label{fig:node}
\end{figure}

Figure~\ref{fig:node} shows the nodal structures, that is, the momentum points at which the dispersion of the wallpaper fermions remains gapless even in the superconducting state.
We plot the momentum points where the energy is smaller than $10^{-4}$ as gapless points.
We find point nodes for $\Delta_2$ and line nodes for $\Delta_5$ and $\Delta_6$ in addition to full-gap structures for $\Delta_1$, $\Delta_3$, and $\Delta_4$.
In the following subsections, we discuss the theoretical mechanisms that give rise to these nodal structures for $\Delta_2$, $\Delta_5$, and $\Delta_6$.

\subsection{Zero-dimensional topological invariant}\label{subsec3-2}
In this subsection, we focus on the BdG Hamiltonian at each point in momentum space and consider the zero-dimensional (0D) symmetry class.
We first formulate a general framework for determining the 0D symmetry class of the BdG Hamiltonian, and then apply it to each momentum region.

Our BdG Hamiltonian~\eqref{eq:BdG} satisfies the following symmetries:
\begin{itemize}
    \item Time-reversal symmetry (TRS)
        \begin{equation}
            \tilde{\Theta}H^{(\mathrm{eff})}_\mathrm{scwp}(\bm{k})\tilde{\Theta}^{-1}=H^{(\mathrm{eff})}_\mathrm{scwp}(-\bm{k}),\quad\tilde{\Theta}\Def-is_\mathrm{y}\sigma_0\tau_0\mathcal{K},
            \label{eq:TRS}
        \end{equation}
    \item Particle-hole symmetry (PHS)
        \begin{equation}
            \tilde{\Xi}H^{(\mathrm{eff})}_\mathrm{scwp}(\bm{k})\tilde{\Xi}^{-1}=-H^{(\mathrm{eff})}_\mathrm{scwp}(-\bm{k}),\quad\tilde{\Xi}\Def-is_\mathrm{y}\sigma_0\tau_\mathrm{y}\mathcal{K},
            \label{eq:PHS}
        \end{equation}
    \item Chiral symmetry (CS)
        \begin{equation}
            \tilde{\Gamma}H^{(\mathrm{eff})}_\mathrm{scwp}(\bm{k})\tilde{\Gamma}^\dagger=-H^{(\mathrm{eff})}_\mathrm{scwp}(\bm{k}),\quad\tilde{\Gamma}\Def s_0\sigma_0\tau_\mathrm{y}.
            \label{eq:CS}
        \end{equation}
\end{itemize}
To define 0D symmetry operators of Eqs.~\eqref{eq:TRS}--\eqref{eq:CS}, we first consider a symmetry operation $d$ that maps $\bm{k}$ to $-\bm{k}$, as
\begin{equation}
    D(d)H^{(\mathrm{eff})}_{\mathrm{wp}}(\bm{k})D(d)^\dagger=H^{(\mathrm{eff})}_{\mathrm{wp}}(-\bm{k}).
\end{equation}
We note that the representation matrix $D(d)$ is defined such that $[D(d),\Theta]=0$ as shown in Table~\ref{tab:rep_mat}.
The representation of $d$ for the BdG Hamiltonian is defined as
\begin{equation}
    \tilde{D}(d)\Def
    \mqty(D(d) & 0 \\ 0 & \chi(d)D(d)),
    \label{eq:d_BdG}
\end{equation}
for $D(d)\hat{\Delta}_iD(d)^\dag=\chi(d)\hat{\Delta}_i$ with $\chi(d)=\pm 1$, and satisfies
\begin{equation}
    \tilde{D}(d)H^{(\mathrm{eff})}_{\mathrm{scwp}}(\bm{k})\tilde{D}(d)^\dagger=H^{(\mathrm{eff})}_{\mathrm{scwp}}(-\bm{k}).
\end{equation}
Using Eq.~\eqref{eq:d_BdG}, we can define the following 0D time-reversal, particle-hole, and chiral operators:
\begin{itemize}
    \item 0D TRS:
        \begin{equation}
            \tilde{\Theta}_0H^{(\mathrm{eff})}_{\mathrm{scwp}}(\bm{k}){\tilde{\Theta}_0}^{-1}=H^{(\mathrm{eff})}_{\mathrm{scwp}}(\bm{k}),\quad \tilde{\Theta}_0\Def\tilde{D}(d)\tilde{\Theta},
            \label{eq:0D_TRS}
        \end{equation}
    \item 0D PHS:
        \begin{equation}
            \tilde{\Xi}_0H^{(\mathrm{eff})}_{\mathrm{scwp}}(\bm{k}){\tilde{\Xi}_0}^{-1}=-H^{(\mathrm{eff})}_{\mathrm{scwp}}(\bm{k}),\quad \tilde{\Xi}_0\Def\tilde{D}(d)\tilde{\Xi},
            \label{eq:0D_PHS}
        \end{equation}
    \item 0D CS:
        \begin{equation}
            \tilde{\Gamma}_0H^{(\mathrm{eff})}_\mathrm{scwp}(\bm{k})\tilde{\Gamma}_0^\dagger=-H^{(\mathrm{eff})}_\mathrm{scwp}(\bm{k}),\quad \tilde{\Gamma}_0\Def\tilde{\Gamma}.
            \label{eq:0D_CS}
        \end{equation}
\end{itemize}

\begin{figure}[htbp]
    \centering
    \begin{minipage}[t]{0.9\hsize}
        \centering
        \includegraphics[keepaspectratio,scale=0.25]{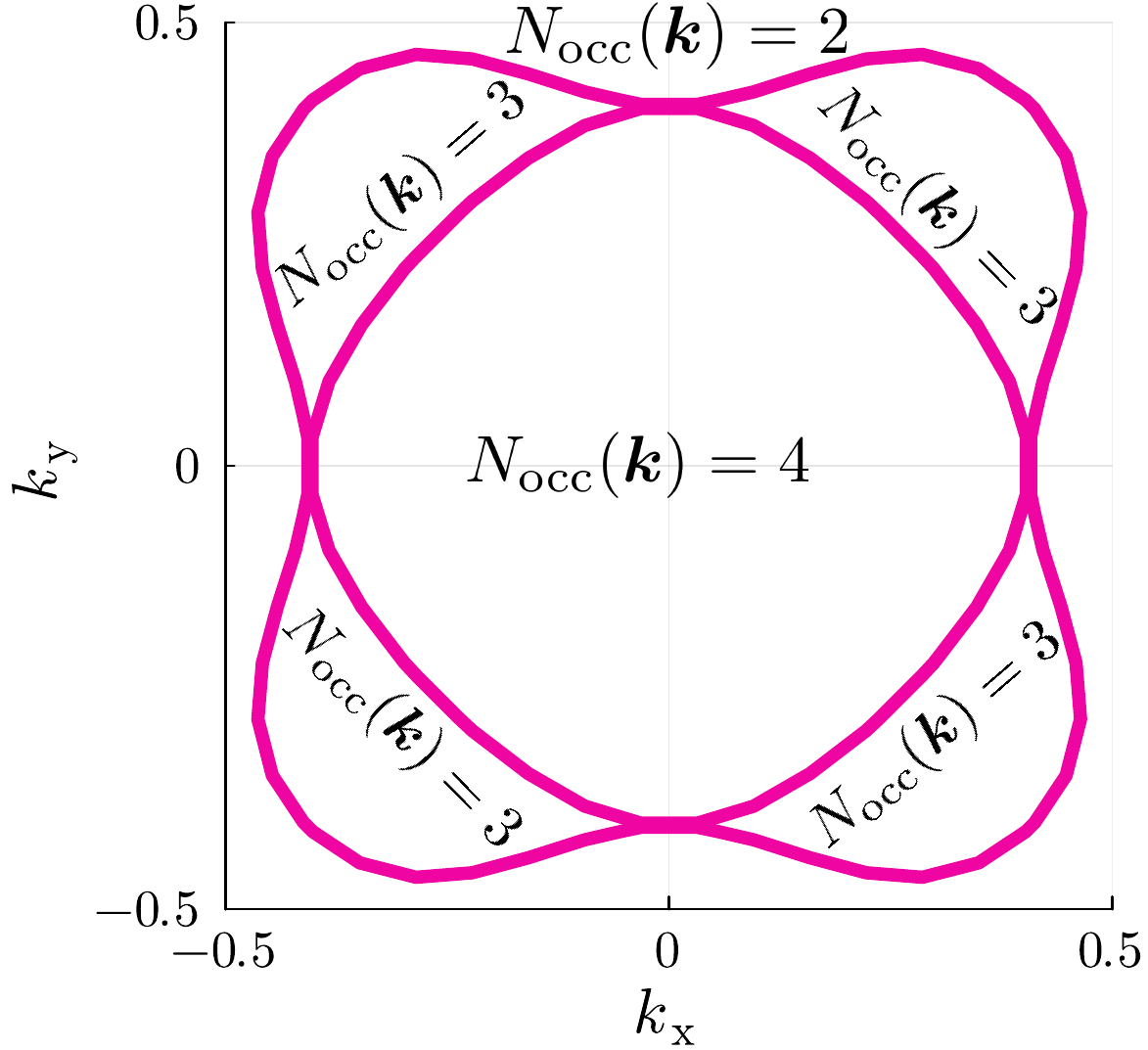}
        \caption{(Color online)
                Fermi surface and the number of occupied bands at each momentum points $N_\mathrm{occ}(\bm{k})$.
                }
        \label{fig:fermi_surface}
    \end{minipage} \\
    \nextfloat
    \begin{minipage}[t]{0.9\hsize}
        \centering
        \begin{minipage}[t]{0.45\hsize}
            \centering
            \captionsetup{position=top, justification=raggedright, singlelinecheck=false}
            \subcaption{}
            \includegraphics[keepaspectratio, scale=0.68]{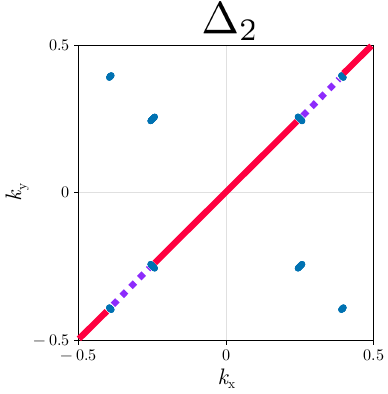}
            \label{fig:nu_delta_2_110}
        \end{minipage}
        \begin{minipage}[t]{0.45\hsize}
            \centering
            \captionsetup{position=top, justification=raggedright, singlelinecheck=false}
            \subcaption{}
            \includegraphics[keepaspectratio, scale=0.68]{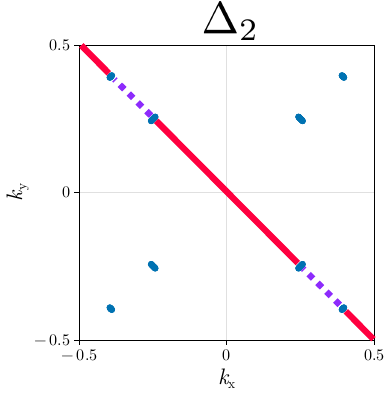}
            \label{fig:nu_delta_2_1-10}
        \end{minipage}
        \caption{(Color online)
                0D topological invariants for the $\Delta_2$ pairing.
                The red solid (purple dashed) line indicates $\nu_{\bm{k}}[d]=0\ (1)$: (a) along the $[110]$ line for $d=\{m_{11}|1/2\ 1/2\}$, and (b) along the $[1\bar{1}0]$ line for $d=\{m_{1\bar{1}}|1/2\ 1/2\}$.
                }
        \label{fig:nu_delta_2}
    \end{minipage} \\
    \nextfloat
    \begin{minipage}[t]{0.9\hsize}
        \centering
        \begin{minipage}[t]{0.45\hsize}
            \centering
            \captionsetup{position=top, justification=raggedright, singlelinecheck=false}
            \subcaption{}
            \includegraphics[keepaspectratio, scale=0.68]{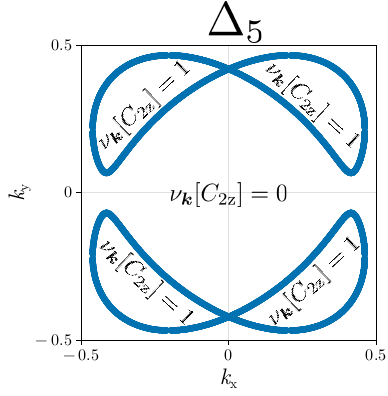}
            \label{fig:nu_delta_5}
        \end{minipage}
        \begin{minipage}[t]{0.45\hsize}
            \centering
            \captionsetup{position=top, justification=raggedright, singlelinecheck=false}
            \subcaption{}
            \includegraphics[keepaspectratio, scale=0.68]{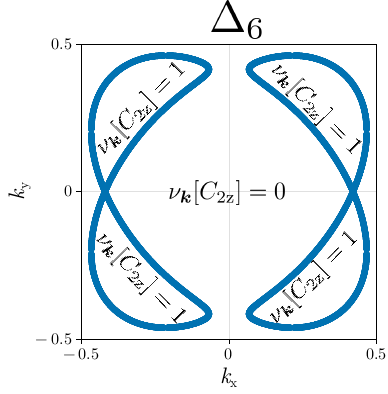}
            \label{fig:nu_delta_6}
        \end{minipage}
        \caption{(Color online)
                0D topological invariants $\nu_{\bm{k}}[C_{2\mathrm{z}}]$ for the (a) $\Delta_5$ and (b) $\Delta_6$ pairings.
                }
        \label{fig:nu_delta_5,6}
    \end{minipage}
\end{figure}

Based on the above construction, we can classify the symmetry classes in zero dimensions.
Our BdG Hamiltonian possesses chiral symmetry for any $\bm{k}$.
$(\tilde{\Theta}_0)^2$ is determined solely by the space group and the $\bm{k}$ point.
Due to $[\tilde{D}(d),\tilde{\Theta}]=0$ and $\tilde{\Theta}^2=-1$, we obtain
\begin{equation}
    (\tilde{\Theta}_0)^2=-D(d)^2\tau_0=\pm 1.
\end{equation}
$(\tilde{\Xi}_0)^2$ represents the parity of the pair potential for $d$, $\chi(d)$, as
\begin{equation}
    (\tilde{\Xi}_0)^2=\chi(d)D(d)^2\tau_0=\pm 1,
\end{equation}
where we use $\tilde{\Xi}^2=+1$.
The 0D symmetry classes under $d$ are summarized in Table~\ref{tab:0D_class}.
\begin{table*}[t]
    \centering
    \caption{
        Symmetry classes and 0D topological invariants.
        $\chi(d)$ and $D(d)^2$ denote the parity of the pair potential for $d$ and the square of representation matrix, respectively.
        The squares ($\pm 1$) of $\tilde{\Theta}_0$, $\tilde{\Xi}_0$, and $\tilde\Gamma_0$ are shown.}
    \begin{tabular}{cccccccc}
        \hline\hline
        $\chi(d)$ & $D(d)^2$ & $\tilde{\Theta}_0$ & $\tilde{\Xi}_0$ & $\tilde{\Gamma}_0$ & Class & Topological invariant \\ \hline
        $-1$ & $-1$ & $+1$ & $+1$ & $1$ & BDI &  $\nu_{\bm{k}}[d]\in\mathbb{Z}_2=\{0,1\}$ \\
        $+1$ & $-1$ & $+1$ & $-1$ & $1$ & CI & 0 \\
        $+1$ & $+1$ & $-1$ & $+1$ & $1$ & DIII & 0 \\
        $-1$ & $+1$ & $-1$ & $-1$ & $1$ & CII & 0 \\
        \hline\hline
    \end{tabular}
    \label{tab:0D_class}
\end{table*}
If $\chi(d)=-1$ and $D(d)^2=-1$, our BdG Hamiltonian at each momentum point belongs to class BDI, which hosts a $\Z_2$ topological invariant $\nu_{\bm{k}}[d]$.
If not, it belongs to class CI, DIII, or CII, for which no topological invariant can be defined.
In the weak-coupling limit, the 0D topological invariant in class BDI is given by~\cite{shiozaki2019variants, geier2020symmetry}
\begin{equation}
    \nu_{\bm{k}}[d]=N_\mathrm{occ}(\bm{k}) \pmod 2,
    \label{eq:nu_def}
\end{equation}
where $N_\mathrm{occ}(\bm{k})$ represents the number of occupied bands of the normal state at momentum $\bm{k}$.
Here, we refer to the parameter range as ``weak-coupling limit'' when $H^{(\mathrm{eff})}_\mathrm{scwp}(\bm{k})$ can be continuously deformed to $H^{(\mathrm{eff})}_\mathrm{scwp}(\bm{k})|_{\Delta_0=0}$.
Figure~\ref{fig:fermi_surface} shows the Fermi surface and $N_\mathrm{occ}(\bm{k})$ in our system.
Taking modulo 2 of these values for each momentum point, we obtain the topological invariant $\nu_{\bm{k}}[d]$.
And, nodes appear at the boundary between regions of two $\Z_2$ phases.

On the $\bar{\mathrm{M}}$ point, the squares of order-two symmetry operations are given by
\begin{align}
    & D(\{2|\bm{0}\})^2=-1, \\
    & D(\{m_{10}|1/2\ 1/2\})^2=D(\{m_{01}|1/2\ 1/2\})^2=+1, \\
    & D(\{m_{11}|1/2\ 1/2\})^2=D(\{m_{1\bar{1}}|1/2\ 1/2\})^2=-1.
\end{align}
Along the $\ev{110}$ lines, $d$ is chosen to be $m_\mathrm{d}\Def\{m_{11}|1/2\ 1/2\}$ and $\{m_{1\bar{1}}|1/2\ 1/2\}$.
The point nodes for $\Delta_2$ pairing, which is odd under $m_\mathrm{d}$, $\chi(m_\mathrm{d})=-1$, are protected by the topological invariant $\nu_{\bm{k}} [m_\mathrm{d}]$, as illustrated in Fig.~\ref{fig:nu_delta_2}.
On one hand, on general points, $d$ is $C_{2\mathrm{z}}\Def\{2|\bm{0}\}$ then $C_{2\mathrm{z}}$-odd representations, $\chi(C_{2\mathrm{z}})=-1$, $\Delta_5$ and $\Delta_6$ remains gapless, resulting in the line nodes, which is protected by $\nu_{\bm{k}}[C_{2\mathrm{z}}]$, as shown in Fig.~\ref{fig:nu_delta_5,6}.

However, not all nodes in these states are captured by this invariant.
Along the $[010]$ line in $\Delta_5$ and the $[100]$ line in $\Delta_6$, nodes appear, even though $\nu_{\bm{k}}[C_{2\mathrm{z}}]$ does not change.
In the case of $\Delta_6$ along the $[100]$ line, the glide operation $m_\mathrm{x}\Def\{m_{10}|1/2\ 1/2\}$ is eligible for $d$.
However, because $\hat{\Delta}_6$ is even parity under $m_\mathrm{x}$ and $D(m_\mathrm{x})^2=1$, the BdG Hamiltonian for $\Delta_6$ on the [100] line belongs to class DIII.
Thus, we cannot define the $\Z_2$ topological invariant $\nu_{\bm{k}}[m_\mathrm{x}]$ for the $\Delta_6$ pairing.
By the same procedure, the topological invariant with respect to $d=\{m_{01}|1/2\ 1/2\}$ cannot be defined on the $[010]$ line for the $\Delta_5$ pairing.
These cases require a separate consideration.

\subsection{Group-theoretical approach}\label{subsec3-3}
We employ the group-theoretical approach~\cite{kobayashi2018symmetryprotected, sumita2018unconventional, sumita2019classification, yoshida2019efficient, yamazaki2021magnetic} to determine whether a superconducting gap opens or remains nodal.
The following assumptions are made (see Fig.~\ref{fig:MB_assumption}):
\begin{figure}[t]
    \centering
    \includegraphics[keepaspectratio,scale=0.5]{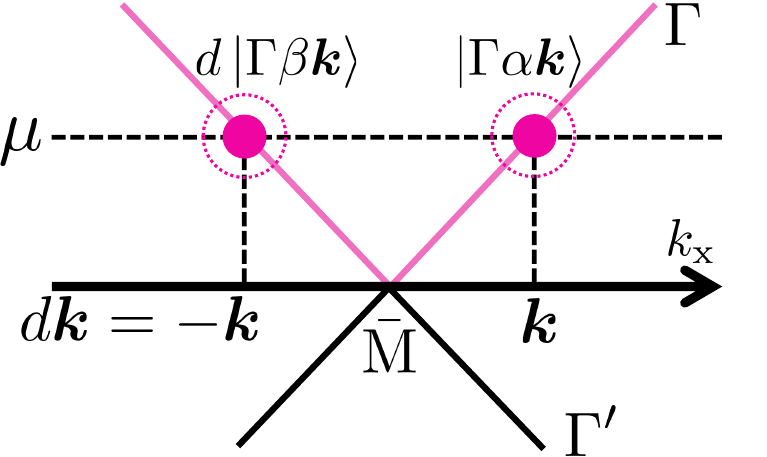}
    \caption{(Color online)
            We consider weak-coupling BCS-type pairing.
            Cooper pairs are assumed to be formed within the same band on the Fermi surface without a finite center-of-mass momentum.
            }
    \label{fig:MB_assumption}
\end{figure}
\begin{itemize}
    \item
        We assume the weak-coupling limit, that is, Cooper pairs are formed on the Fermi surface, and no inter-band pairing is considered.
    \item
        The center-of-mass momentum of the Cooper pairs is set to zero.
        Namely, two Bloch states at momenta $\bm{k}$ and $d\bm{k}=-\bm{k}$ form a Cooper pair.
\end{itemize}
We discuss the intra-band order parameter under this assumption.
The BdG Hamiltonian can be rewritten in the band basis, which diagonalizes the normal Hamiltonian.
With $\tilde{U}(\bm{k})\Def U(\bm{k})\tau_0$, where $U(\bm{k})H^{(\mathrm{eff})}_{\mathrm{wp}}(\bm{k})U(\bm{k})^\dagger=\mathrm{diag}[\varepsilon_1(\bm{k}),\varepsilon_2(\bm{k}),\varepsilon_3(\bm{k}),\varepsilon_4(\bm{k})]\deF\varepsilon(\bm{k})$, the BdG Hamiltonian takes the form
\begin{equation}
    H_\mathrm{Band}(\bm{k})\Def
    \tilde{U}(\bm{k})H^{(\mathrm{eff})}_\mathrm{scwp}(\bm{k})\tilde{U}(\bm{k})^\dagger=
    \mqty(\varepsilon(\bm{k})-\mu & \hat{\Delta}_{\mathrm{Band}}(\bm{k}) \\ \hat{\Delta}_{\mathrm{Band}}(\bm{k}) & -\varepsilon(\bm{k})+\mu),
    \label{eq:BdG_band}
\end{equation}
where
\begin{equation}
    \hat{\Delta}_\mathrm{Band}(\bm{k})\Def
    U(\bm{k})\hat{\Delta} U(\bm{k})^\dagger \sim
    \mqty(\hat{\Delta}^{\Gamma\Gamma}(\bm{k}) & 0 \\ 0 & \hat{\Delta}^{\Gamma'\Gamma'}(\bm{k}))
    \label{eq:Delta_band}
\end{equation}
and $\Gamma$ and $\Gamma'$ denote the band representations.
Since the off-diagonal component $\hat{\Delta}^{\Gamma\Gamma'}(\bm{k})$ (inter-band order parameter) in Eq.~\eqref{eq:Delta_band} contributes to the superconducting gap only at third order, we therefore neglect this small effect.
$(\hat{\Delta}^{\Gamma\Gamma})_{\alpha,\beta}(\bm{k})\sim\ket{\Gamma\alpha\bm{k}}\otimes d\ket{\Gamma\beta\bm{k}}$ represents the intra-band order parameter.
Here, the Bloch states form a representation basis of the small co-representation $\gamma_{\bm{k}}$ of the magnetic little group $M_{\bm{k}}\Def G_{\bm{k}}+G_{\bm{k}}d\Theta$, where $G_{\bm{k}}$ is the little group of the space group $G$ and $\Theta$ is the time-reversal operator.
The intra-band order parameter $\hat{\Delta}^{\Gamma\Gamma}(\bm{k})$ follows the antisymmetrized representation $P_{\bm{k}}$ spanned by $\ket{\Gamma\alpha\bm{k}}\otimes d\ket{\Gamma\beta\bm{k}}$.
The character $\chi[P_{\bm{k}}(m)]$ for the representation $P_{\bm{k}}$ of the magnetic space group $\mathcal{M}_{\bm{k}}\Def M_{\bm{k}}+dM_{\bm{k}}$ can be calculated from the character $\chi[\gamma_{\bm{k}}(m)]$ for the small co-representation $\gamma_{\bm{k}}$ by using the Mackey--Bradley theorem~\cite{mackey1953symmetric, bradley1970kronecker, bradley2009mathematical},
\begin{subequations}
    \begin{align}
        &\chi[P_{\bm{k}}(m)]=\chi[\gamma_{\bm{k}}(m)]\chi[\gamma_{\bm{k}}(d^{-1}md)], \\
        &\chi[P_{\bm{k}}(dm)]=-\chi[\gamma_{\bm{k}}(dmdm)],
    \end{align}
    \label{eq:MB}
\end{subequations}
where $m\in G_{\bm{k}}$.
Then we decompose the $P_{\bm{k}}$ into irreps of the group $\mathcal{G}_{\bm{k}}\Def G_{\bm{k}}+dG_{\bm{k}}$.
Since the pair potential is invariant under translations, it suffices to consider the decomposition into the irreps of the point group $\mathcal{G}_{\bm{k}}/T$, where $T$ denotes the translation group.
If a given irrep of pair potential is excluded in this decomposition, no superconducting gap opens.
Note that the point group $\mathcal{G}_{\bm{k}}/T$ is independent of the choice of $d$~\cite{sumita2018unconventional}.
In the following, we fix $d=\{2|\bm{0}\}$.
The symmetry operations that appear on the right-hand side of Eq.~\eqref{eq:MB} are summarized in Table~\ref{tab:symm_ope_MB}.
\begin{table*}[t]
    \centering
    \caption{
        Symmetry operations in Mackey--Bradley theorem for wallpaper group $\mathrm{p4g}$. We set $d=\{2|\bm{0}\}\deF C_{2\mathrm{z}}$.
        The superscript $d$ in $\{{}^dp_g|\bm{\tau}_g\}$ indicates that the operation corresponds to a $2\pi$ rotation of the symmetry operation $\{p_g|\bm{\tau}_g\}$.
        }
    \begin{tabular}{llll}
        \hline\hline
        $m$ & $C_{2\mathrm{z}}m$ & $C_{2\mathrm{z}}^{-1}mC_{2\mathrm{z}}$ & $C_{2\mathrm{z}}mC_{2\mathrm{z}}m$ \\ \hline
        $\{E|\bm{0}\}$ & $\{2|\bm{0}\}$ & $\{E|\bm{0}\}$ & $\{{}^d E|\bm{0}\}$ \\
        $\{2|\bm{0}\}$ & $\{{}^d E|\bm{0}\}$ & $\{2|\bm{0}\}$ & $\{E|\bm{0}\}$ \\
        $\{m_{10}|1/2\ 1/2\}$ & $\{m_{01}|-1/2\ -1/2\}$ & $\{{}^d m_{10}|-1/2\ -1/2\}$ & $\{{}^d E|-1\ 0\}$ \\
        $\{m_{1\bar{1}}|1/2\ 1/2\}$ & $\{m_{11}|-1/2\ -1/2\}$ & $\{{}^d m_{1\bar{1}}|-1/2\ -1/2\}$ & $\{{}^d E|\bm{0}\}$ \\
        \hline\hline
    \end{tabular}
    \label{tab:symm_ope_MB}
\end{table*}

We apply the above discussion to the $[100]$ line.
The little group $G_{\bm{k}}$ along the $[100]$ line is written as
\begin{equation}
    G_{\bm{k}}=\{E|\bm{0}\}T+\{m_{01}|1/2\ 1/2\}T.
\end{equation}
In this case, the point group $\mathcal{G}_{\bm{k}}/T$ generated from the little group $G_{\bm{k}}$ is isomorphic to the point group $\mathrm{C}_{2\mathrm{v}}$.
The representation of twofold degenerate bands is given by $\bar{\mathrm{Y}}_3\bar{\mathrm{Y}}_4(2)$~\cite{Aroyo2011-cr, Aroyo2006-bi1, Aroyo2006-bi2, elcoro2021magnetic, xu2020high}.
The characters are obtained as
\begin{equation}
    \chi[\gamma_{\bm{k}}(\{E|\bm{0}\})]=2,\quad\chi[\gamma_{\bm{k}}(\{m_{01}|1/2\ 1/2\})]=0.
\end{equation}
Substituting these into Eq.~\eqref{eq:MB}, we obtain the characters listed in Table~\ref{tab:MB_100}.
\begin{table*}[t]
    \centering
    \caption{
            The characters for representations $P_{\bm{k}}$ of the pair potentials on (a) $[100]$ and (b) $[110]$ line, and (c) general momentum points.
            The point group $\mathcal{G}_{\bm{k}}/T$ is $\mathrm{C_{2v}}$ for (a) and (b), and $\mathrm{C_2}$ for (c).
            The column labeled ``Decomposition'' shows the result of decomposing $P_{\bm{k}}$ into the irreps of $\mathcal{G}_{\bm{k}}/T$.
            This decomposition is compatible with the decomposition into the irreps of $\mathrm{C}_{4\mathrm{v}}$ shown in the rightmost column.
            Pair potentials corresponding to the representations not contained in this irreducible decomposition give rise to nodes.
            }
    \begin{minipage}[t]{1.0\hsize}
        \captionsetup{position=top, justification=raggedright, singlelinecheck=false}
        \subcaption{$[100]$ line.}
        \label{tab:MB_100}
        \begin{tabular}{lcccccl}
            \hline\hline
            $\mathrm{C}_{2\mathrm{v}}$ & $E$ & $C_{2\mathrm{z}}$ & $m_{01}$ & $m_{10}$ & Decomposition & Compatible in $\mathrm{C}_{4\mathrm{v}}$\\ \hline
            $P_{[100]}$ & $4$ & $2$ & $0$ & $-2$ & $\mathrm{A}_1\oplus2\mathrm{A}_2\oplus\mathrm{B}_1$ & $\mathrm{A}_1\oplus2\mathrm{A}_2\oplus\mathrm{B}_1\oplus2\mathrm{B}_2\oplus{}^1\mathrm{E}$ \\
            \hline\hline
        \end{tabular}
    \end{minipage} \\
    \begin{minipage}[t]{1.0\hsize}
        \captionsetup{position=top, justification=raggedright, singlelinecheck=false}
        \subcaption{$[110]$ line.}
        \label{tab:MB_110}
        \begin{tabular}{lcccccl}
            \hline\hline
            $\mathrm{C}_{2\mathrm{v}}$ & $E$ & $C_{2\mathrm{z}}$ & $m_{1\bar{1}}$ & $m_{11}$ & Decomposition & Compatible in $\mathrm{C}_{4\mathrm{v}}$ \\ \hline
            $P_{[110]}$ & $1$ & $1$ & $1$ & $1$ & $\mathrm{A}_1$ & $\mathrm{A}_1\oplus\mathrm{B}_2$\\
            \hline\hline
        \end{tabular}
    \end{minipage} \\
    \begin{minipage}[t]{1.0\hsize}
        \captionsetup{position=top, justification=raggedright, singlelinecheck=false}
        \subcaption{General points.}
        \label{tab:MB_GP}
        \begin{tabular}{lcccl}
            \hline\hline
                $\mathrm{C}_2$ & $E$ & $C_{2\mathrm{z}}$ & Decomposition & Compatible in $\mathrm{C}_{4\mathrm{v}}$ \\ \hline
                $P_\mathrm{GP}$ & $1$ & $1$ & $\mathrm{A}$ & $\mathrm{A}_1\oplus\mathrm{A}_2\oplus\mathrm{B}_1\oplus\mathrm{B}_2$\\
            \hline\hline
        \end{tabular}
    \end{minipage}
\end{table*}
Here, we have used the invariance of the representation of the pair potential under translation, $P_{\bm{k}}(mt)=P_{\bm{k}}(m)$ with $t\in T$.
The decomposition into the irreps of the point group $\mathrm{C}_{2\mathrm{v}}$ yields $\mathrm{A}_1\oplus2\mathrm{A}_2\oplus\mathrm{B}_1$.
From the compatibility relations, this decomposition corresponds to
\begin{equation}
    \mathrm{A}_1\oplus2\mathrm{A}_2\oplus\mathrm{B}_1\oplus2\mathrm{B}_2\oplus{}^1\mathrm{E}
\end{equation}
in the irreps of the point group $\mathrm{C}_{4\mathrm{v}}$.
The ${}^2\mathrm{E}$ representation, which corresponds to $\Delta_6$, is not included in this decomposition and should vanish along the $[100]$ line.
Consequently, $\Delta_6$ forms a node on the $[100]$ line.
$\Delta_5$ is obtained from $\Delta_6$ by the fourfold rotation, and the node appears on the $[010]$ line.
From the discussion in Secs.~\ref{subsec3-2} and \ref{subsec3-3}, the nodal structures in Fig.~\ref{fig:node} are theoretically justified.

The Mackey--Bradley theorem provides a comprehensive framework for dealing with crystalline symmetries and the presence or absence of a superconducting gap.
Since the point nodes of the $\Delta_2$ pair potential and the line nodes of $\Delta_5$ (except along the $[010]$ line) and $\Delta_6$ (except along the $[100]$ line) pair potentials discussed in Sec.~\ref{subsec3-2} are characterized by the $\Z_2$ topological invariants defined by a crystalline symmetry operator $d$, they should also be explained consistently within the Mackey--Bradley theorem.
Tables~\ref{tab:MB_110} and \ref{tab:MB_GP} summarize the results obtained by applying the Mackey–Bradley theorem to the $[110]$ line and the general momentum points, respectively.
From these results, one finds that the discussion based on the 0D topological invariant is consistent with that based on the Mackey--Bradley theorem.
The details of the calculations are provided in Appendix~\ref{app1}.

\section{Discussion}\label{sec4}
To discuss the differences between the cases of nodes protected by topological invariants, as in Sec.~\ref{subsec3-2}, and those protected by crystalline symmetries, as in Sec.~\ref{subsec3-3}, we consider the situation where $\Delta_0$ is large.
We discuss the case of topologically protected nodes.
The topological invariant can be defined as follows, without restriction to the weak-coupling regime~\cite{shiozaki2019variants, geier2020symmetry, ono2021Z2, ono2022symmetrybased}
\begin{equation}
    \nu_{\bm{k}}[d]\Def\frac{1}{i\pi}\log\frac{\Pf\qty[\tilde{\mathcal{C}}_0[d]H^{(\mathrm{eff})}_\mathrm{scwp}(\bm{k})]}{\Pf\qty[\tilde{\mathcal{C}}_0[d]H^{(\mathrm{eff})}_\mathrm{scwp}(\bm{k}_0)]} \pmod 2,
\end{equation}
where $\tilde{\mathcal{C}}_0[d]$ is a unitary part of $\tilde{\Xi}_0$ and $\bm{k}_0$ denotes a reference point.
Accordingly, the nodes protected by the topological invariant discussed in the previous section remain stable even at a finite magnitude of pair potential.

\section{Conclusion}\label{sec5}
In this work, we investigated the superconducting gap structures of wallpaper fermions within an effective model framework.
Our numerical analysis demonstrated that among the six momentum-independent pair potentials allowed by crystalline symmetry, $\Delta_1$, $\Delta_3$, and $\Delta_4$ lead to fully gapped structures, while $\Delta_2$ hosts point nodes and $\Delta_5$ and $\Delta_6$ exhibit line nodes.
Using topological invariants and group-theoretical approaches, we clarified the distinct protection mechanisms of these nodal structures.
The point nodes in $\Delta_2$ pairing and most of the line nodes in $\Delta_5$ and $\Delta_6$ are protected by 0D topological invariants.
In contrast, the nodes along the $[010]$ line for $\Delta_5$ and along the $[100]$ line for $\Delta_6$ are protected by crystalline symmetry through the Mackey--Bradley theorem.
These findings reveal that in the superconducting state, the wallpaper fermions can remain gapless states enforced by nonsymmorphic wallpaper group $\mathrm{p4g}$ symmetry.

\backmatter

\bmhead{Acknowledgements}
K.Y. is supported by Murata Science and Education Foundation (Grants No.~M25SN042) and Marubun Research Promotion Foundation.
A.Y. is supported by JSPS KAKENHI for Grants (Grants Nos.~JP24H00853 and JP25K07224).

\section*{Declarations}
\begin{itemize}
    \item Funding \\
        K.Y. is supported by Murata Science and Education Foundation (Grants No.~M25SN042) and Marubun Research Promotion Foundation.
        A.Y. is supported by JSPS KAKENHI for Grants (Grants Nos.~JP24H00853 and JP25K07224).
    \item Data availability \\
        The data supporting the findings of this study are available at NAGOYA Repository (URL: \url{https://nagoya.repo.nii.ac.jp/?page=1&size=20&sort=controlnumber&search_type=0&q=0}).
\end{itemize}

\begin{appendices}

\section{Detailed calculations}\label{app1}

\subsection{\texorpdfstring{$[110]$}{[110]} line}\label{app1-1}
Along the $[110]$ line, $\bm{k} = (k,k)$, the little group is
\begin{equation}
    G_{\bm{k}}=\{E|\bm{0}\}T+\{m_{1\bar{1}}|1/2\ 1/2\}T.
\end{equation}
The four-dimensional representation of the wallpaper fermion on the $\bar{\mathrm{M}}$ point is split into $\bar{\Sigma}_3(1)\oplus\bar{\Sigma}_4(1)$ on the $[110]$ line~\cite{Aroyo2011-cr, Aroyo2006-bi1, Aroyo2006-bi2, elcoro2021magnetic, xu2020high}, whose characters are
\begin{subequations}
    \begin{align}
    &\chi[\gamma_{\bm{k}}(\{E|t_1\;t_2\})]=e^{2i\pi(t_1+t_2)k},
    \end{align}
    \begin{empheq}[left={\chi[\gamma_{\bm{k}}(\{m_{1\bar{1}}|1/2\ 1/2\})]=\empheqlbrace}]{alignat = 2}
        & -ie^{2i\pi k} & \quad & \mathrm{for}\ \bar{\Sigma}_3, \\
        & ie^{2i\pi k} & \quad & \mathrm{for}\ \bar{\Sigma}_4.
    \end{empheq}
\end{subequations}
Applying Eq.~\eqref{eq:MB} in the same manner as in Sec.~\ref{subsec3-3}, one obtains the characters summarized in Table~\ref{tab:MB_110}.
Decomposing into irreps of $\mathcal{G}_{\bm{k}}/T\simeq\mathrm{C}_{2\mathrm{v}}$, and using the compatibility relation, we obtain
\begin{equation}
    \mathrm{A}_1\oplus\mathrm{B}_2.
\end{equation}
Thus, the representations $\mathrm{A}_2$ and $\mathrm{E}$ are excluded, indicating that $\Delta_2$, $\Delta_5$ and $\Delta_6$ host nodes along the $[110]$ line.

\subsection{General points}\label{app1-2}
For general momentum points (excluding the $[100]$, $[010]$, $[110]$, and $[1\bar{1}0]$ lines), the little group reduces to
\begin{equation}
    G_{\bm{k}}=\{E|\bm{0}\}T,
\end{equation}
with character
\begin{equation}
    \chi[\gamma_{\bm{k}}(\{E|\bm{0}\})]=1.
\end{equation}
The Mackey--Bradley theorem then yields the characters in Table~\ref{tab:MB_GP}.
Decomposing into irreps of $\mathcal{G}_{\bm{k}}/T\simeq\mathrm{C}_2$, and constructing irreducible decomposition into $\mathrm{C}_{4\mathrm{v}}$, we obtain
\begin{equation}
    \mathrm{A}_1\oplus\mathrm{A}_2\oplus\mathrm{B}_1\oplus\mathrm{B}_2.
\end{equation}
Hence, the $\mathrm{E}$ representation is absent, implying that $\Delta_5$ and $\Delta_6$ necessarily form nodes at general momentum points.
\end{appendices}

\bibliography{ref}

@article{hao2011surface,
  title = {Surface Spectral Function in the Superconducting State of a Topological Insulator},
  author = {Hao, Lei and Lee, T. K.},
  year = {2011},
  month = apr,
  journal = {Phys. Rev. B},
  volume = {83},
  numpages = {13},
  pages = {134516},
  publisher = {American Physical Society},
  doi = {10.1103/PhysRevB.83.134516},
  urldate = {2024-09-23}
}

@article{hashimoto2015surface,
  title = {Surface Electronic State of Superconducting Topological Crystalline Insulator},
  author = {Hashimoto, Tatsuki and Yada, Keiji and Sato, Masatoshi and Tanaka, Yukio},
  year = {2015},
  month = nov,
  journal = {Phys. Rev. B},
  volume = {92},
  numpages = {17},
  pages = {174527},
  publisher = {American Physical Society},
  doi = {10.1103/PhysRevB.92.174527},
  urldate = {2024-09-22}
}

@article{hsieh2012Majorana,
  title = {Majorana {Fermions} and {Exotic} {Surface} {Andreev} {Bound} {States} in {Topological} {Superconductors}: {Application} to {Cu}\textsubscript{x}{Bi}\textsubscript{2}{Se}\textsubscript{3}},
  author = {Hsieh, Timothy H. and Fu, Liang},
  journal = {Phys. Rev. Lett.},
  volume = {108},
  issue = {10},
  pages = {107005},
  numpages = {5},
  year = {2012},
  month = {Mar},
  publisher = {American Physical Society},
  doi = {10.1103/PhysRevLett.108.107005}
}

@article{kawakami2018topological,
  title = {Topological {{Crystalline Materials}} of ${J}=3/2$ {{Electrons}}: {{Antiperovskites}}, {{Dirac Points}}, and {{High Winding Topological Superconductivity}}},
  shorttitle = {Topological {{Crystalline Materials}} of {{J}} = 3 / 2 {{Electrons}}},
  author = {Kawakami, Takuto and Okamura, Tetsuya and Kobayashi, Shingo and Sato, Masatoshi},
  year = {2018},
  month = nov,
  journal = {Phys. Rev. X},
  volume = {8},
  numpages = {4},
  pages = {041026},
  issn = {2160-3308},
  doi = {10.1103/PhysRevX.8.041026},
  urldate = {2024-09-13},
  langid = {english}
}

@article{kobayashi2018symmetryprotected,
  title = {Symmetry-Protected Line Nodes and {{Majorana}} Flat Bands in Nodal Crystalline Superconductors},
  author = {Kobayashi, Shingo and Sumita, Shuntaro and Yanase, Youichi and Sato, Masatoshi},
  year = {2018},
  month = may,
  journal = {Phys. Rev. B},
  volume = {97},
  numpages = {18},
  pages = {180504},
  issn = {2469-9950, 2469-9969},
  doi = {10.1103/PhysRevB.97.180504},
  urldate = {2024-12-10},
  langid = {english}
}

@article{lifshitzanomalies,
  title = {{{Anomalies of Electron Characteristics of a Metal in the High Pressure Region}}},
  author = {Lifshitz, I. M.},
  year = {1960},
  month = nov,
  journal = {Zh. Eksp. Teor. Fiz.},
  volume = {38},
  number = {5},
  pages = {1569},
  langid = {russian}
}

@article{lu2015crossed,
  title = {Crossed {{Surface Flat Bands}} of {{Weyl Semimetal Superconductors}}},
  author = {Lu, Bo and Yada, Keiji and Sato, Masatoshi and Tanaka, Yukio},
  year = {2015},
  month = mar,
  journal = {Phys. Rev. Lett.},
  volume = {114},
  numpages = {9},
  pages = {096804},
  issn = {0031-9007, 1079-7114},
  doi = {10.1103/PhysRevLett.114.096804},
  urldate = {2025-08-19},
  copyright = {http://link.aps.org/licenses/aps-default-license},
  langid = {english}
}

@article{mizuno2023hall,
  title = {Hall Effect of Ferro/Antiferromagnetic Wallpaper Fermions},
  author = {Mizuno, Koki and Yamakage, Ai},
  year = {2023},
  month = jun,
  journal = {Phys. Rev. B},
  volume = {107},
  numpages = {23},
  pages = {235301},
  publisher = {American Physical Society},
  doi = {10.1103/PhysRevB.107.235301},
  urldate = {2024-09-23}
}

@article{novak2013unusual,
  title = {Unusual Nature of Fully Gapped Superconductivity in {{In-doped SnTe}}},
  author = {Novak, Mario and Sasaki, Satoshi and Kriener, Markus and Segawa, Kouji and Ando, Yoichi},
  year = {2013},
  month = oct,
  journal = {Phys. Rev. B},
  volume = {88},
  numpages = {14},
  pages = {140502},
  issn = {1098-0121, 1550-235X},
  doi = {10.1103/PhysRevB.88.140502},
  urldate = {2025-08-19},
  copyright = {http://link.aps.org/licenses/aps-default-license},
  langid = {english}
}

@article{ono2021Z2,
  title = {$\mathbb{Z}_2$-enriched symmetry indicators for topological superconductors in the 1651 magnetic space groups},
  author = {Ono, Seishiro and Po, Hoi Chun and Shiozaki, Ken},
  year = {2021},
  month = may,
  journal = {Phys. Rev. Res.},
  volume = {3},
  numpages = {2},
  pages = {023086},
  issn = {2643-1564},
  doi = {10.1103/PhysRevResearch.3.023086},
  urldate = {2024-11-21},
  langid = {english}
}

@article{ono2022symmetrybased,
  title = {Symmetry-{{Based Approach}} to {{Superconducting Nodes}}: {{Unification}} of {{Compatibility Conditions}} and {{Gapless Point Classifications}}},
  shorttitle = {Symmetry-{{Based Approach}} to {{Superconducting Nodes}}},
  author = {Ono, Seishiro and Shiozaki, Ken},
  year = {2022},
  month = feb,
  journal = {Phys. Rev. X},
  volume = {12},
  numpages = {1},
  pages = {011021},
  publisher = {American Physical Society},
  doi = {10.1103/PhysRevX.12.011021},
  urldate = {2024-11-01}
}

@misc{shiozaki2019variants,
  title = {Variants of the Symmetry-Based Indicator},
  author = {Shiozaki, Ken},
  year = {2019},
  month = jul,
  number = {arXiv:1907.13632},
  eprint = {1907.13632},
  publisher = {arXiv},
  doi = {10.48550/arXiv.1907.13632},
  urldate = {2024-10-28},
  archiveprefix = {arXiv}
}

@article{sumita2018unconventional,
  title = {Unconventional Superconducting Gap Structure Protected by Space Group Symmetry},
  author = {Sumita, Shuntaro and Yanase, Youichi},
  year = {2018},
  month = apr,
  journal = {Phys. Rev. B},
  volume = {97},
  numpages = {13},
  pages = {134512},
  publisher = {American Physical Society},
  doi = {10.1103/PhysRevB.97.134512},
  urldate = {2024-11-12}
}

@article{sumita2019classification,
  title = {Classification of Topological Crystalline Superconducting Nodes on High-Symmetry Lines: {{Point}} Nodes, Line Nodes, and {{Bogoliubov Fermi}} Surfaces},
  shorttitle = {Classification of Topological Crystalline Superconducting Nodes on High-Symmetry Lines},
  author = {Sumita, Shuntaro and Nomoto, Takuya and Shiozaki, Ken and Yanase, Youichi},
  year = {2019},
  month = apr,
  journal = {Phys. Rev. B},
  volume = {99},
  numpages = {13},
  pages = {134513},
  publisher = {American Physical Society},
  doi = {10.1103/PhysRevB.99.134513},
  urldate = {2024-10-25}
}

@article{wieder2018wallpaper,
  title = {Wallpaper Fermions and the Nonsymmorphic {{Dirac}} Insulator},
  author = {Wieder, Benjamin J. and Bradlyn, Barry and Wang, Zhijun and Cano, Jennifer and Kim, Youngkuk and Kim, Hyeong-Seok D. and Rappe, Andrew M. and Kane, C. L. and Bernevig, B. Andrei},
  year = {2018},
  month = jul,
  journal = {Science},
  volume = {361},
  number = {6399},
  pages = {246--251},
  publisher = {American Association for the Advancement of Science},
  doi = {10.1126/science.aan2802},
  urldate = {2025-06-02}
}

@article{yamakage2012theory,
  title = {Theory of Tunneling Conductance and Surface-State Transition in Superconducting Topological Insulators},
  author = {Yamakage, Ai and Yada, Keiji and Sato, Masatoshi and Tanaka, Yukio},
  year = {2012},
  month = may,
  journal = {Phys. Rev. B},
  volume = {85},
  numpages = {18},
  pages = {180509},
  issn = {1098-0121, 1550-235X},
  doi = {10.1103/PhysRevB.85.180509},
  urldate = {2024-09-13},
  copyright = {http://link.aps.org/licenses/aps-default-license},
  langid = {english}
}

@article{yamakage2013anomalous,
  title = {Anomalous {{Josephson}} Current in Superconducting Topological Insulator},
  author = {Yamakage, Ai and Sato, Masatoshi and Yada, Keiji and Kashiwaya, Satoshi and Tanaka, Yukio},
  year = {2013},
  month = mar,
  journal = {Phys. Rev. B},
  volume = {87},
  numpages = {10},
  pages = {100510},
  publisher = {American Physical Society},
  doi = {10.1103/PhysRevB.87.100510},
  urldate = {2025-01-07}
}

@article{yamakage2013theory,
  title = {Theory of Tunneling Spectroscopy in a Superconducting Topological Insulator},
  author = {Yamakage, Ai and Yada, Keiji and Sato, Masatoshi and Tanaka, Yukio},
  year = {2013},
  month = nov,
  journal = {Physica C: Superconductivity},
  volume = {494},
  pages = {20--23},
  issn = {09214534},
  doi = {10.1016/j.physc.2013.04.019},
  urldate = {2024-10-02},
  langid = {english}
}

@article{yamazaki2021magnetic,
  title = {Magnetic Response of {{Majorana Kramers}} Pairs with an Order-Two Symmetry},
  author = {Yamazaki, Yuki and Kobayashi, Shingo and Yamakage, Ai},
  year = {2021},
  month = mar,
  journal = {Phys. Rev. B},
  volume = {103},
  numpages = {9},
  pages = {094508},
  publisher = {American Physical Society},
  doi = {10.1103/PhysRevB.103.094508},
  urldate = {2024-11-22}
}

@article{yoshida2019efficient,
  title = {Efficient Method to Compute $\mathbb{Z}_4$ Indices with Glide Symmetry and Applications to the {{M}}\"{o}bius Materials {{CeNiSn}} and {{UCoGe}}},
  author = {Yoshida, Tsuneya and Daido, Akito and Kawakami, Norio and Yanase, Youichi},
  year = {2019},
  month = jun,
  journal = {Phys. Rev. B},
  volume = {99},
  numpages = {23},
  pages = {235105},
  publisher = {American Physical Society},
  doi = {10.1103/PhysRevB.99.235105},
  urldate = {2024-10-21}
}

@article{mackey1953symmetric,
  author = {George W. Mackey},
  journal = {Am. J. Math.},
  number = {2},
  pages = {387},
  title = {Symmetric and {Anti} {Symmetric} {Kronecker} {Squares} and {Intertwining} {Numbers} of {Induced} {Representations} of {Finite} {Groups}},
  urldate = {2025-08-19},
  volume = {75},
  year = {1953},
  doi = {10.2307/2372459}
}

@article{bradley1970kronecker,
  author = {Bradley, C. J. and Davies, B. L.},
  journal = {J. Math. Phys.},
  number = {5},
  pages = {1536},
  title = {Kronecker {Products} and {Symmetrized} {Squares} of {Irreducible} {Representations} of {Space} {Groups}},
  urldate = {2025-08-19},
  volume = {11},
  year = {1970},
  doi = {10.1063/1.1665292}
}

@article{zhou2021glide,
  title = {Glide Symmetry Protected Higher-Order Topological Insulators from Semimetals with Butterfly-like Nodal Lines},
  author = {Zhou, Xiaoting and Hsu, Chuang-Han and Huang, Cheng-Yi and Iraola, Mikel and Ma{\~n}es, Juan L. and Vergniory, Maia G. and Lin, Hsin and Kioussis, Nicholas},
  year = {2021},
  month = dec,
  journal = {npj Comput. Mater.},
  volume = {7},
  number = {1},
  pages = {202},
  publisher = {Nature Publishing Group},
  issn = {2057-3960},
  doi = {10.1038/s41524-021-00672-9},
  urldate = {2025-08-19},
  copyright = {2021 The Author(s)},
  langid = {english}
}

@article{hwang2023magnetic,
  title = {Magnetic Wallpaper {{Dirac}} Fermions and Topological Magnetic {{Dirac}} Insulators},
  author = {Hwang, Yoonseok and Qian, Yuting and Kang, Junha and Lee, Jehyun and Ryu, Dongchoon and Choi, Hong Chul and Yang, Bohm-Jung},
  year = {2023},
  month = apr,
  journal = {npj Comput. Mater.},
  volume = {9},
  number = {1},
  pages = {65},
  issn = {2057-3960},
  doi = {10.1038/s41524-023-01018-3}
}

@article{ryu2020wallpaper,
  title = {Wallpaper {{Dirac Fermion}} in a {{Nonsymmorphic Topological Kondo Insulator}}: {Pu}{B}${}_4$},
  author = {Ryu, Dong-Choon and Kim, Junwon and Choi, Hongchul and Min, Byung Il},
  year = {2020},
  month = nov,
  journal = {J. Am. Chem. Soc.},
  volume = {142},
  number = {45},
  pages = {19278--19282},
  publisher = {American Chemical Society},
  issn = {0002-7863},
  doi = {10.1021/jacs.0c09442}
}

@article{mizuno2025magnon,
  title = {Magnon-mediated superconductivity at the interface between a ferromagnetic insulator and a topological crystalline insulator with wallpaper fermions},
  author = {Mizuno, Koki and Yamakage, Ai},
  journal = {Phys. Rev. B},
  volume = {112},
  issue = {3},
  pages = {035303},
  numpages = {11},
  year = {2025},
  month = {Jul},
  publisher = {American Physical Society},
  doi = {10.1103/sknv-r7wq}
}

@book{bradley2009mathematical,
  title = {The {Mathematical} {Theory} {Of} {Symmetry} {In} {Solids}: {Representation} theory for point groups and space groups},
  isbn = {978-0-19-958258-7},
  publisher = {Oxford University Press},
  author = {Bradley, C J and Cracknell, A P},
  month = dec,
  year = {2009},
  doi = {10.1093/oso/9780199582587.001.0001},
}

@article{Aroyo2011-cr,
	author = {Aroyo, M.I. and Perez-Mato, J.M. and Orobengoa, D. and Tasci, E. and De La Flor, G. and Kirov, A.},
	title = {Crystallography online: Bilbao crystallographic server},
	year = {2011},
	journal = {Bulgarian Chemical Communications},
	volume = {43},
	number = {2},
	pages = {183 – 197}
}

@article{Aroyo2006-bi1,
  title       = {Bilbao {Crystallographic} {Server}: I. {Databases} and crystallographic computing programs},
  author      = {Mois Ilia Aroyo and Juan Manuel Perez-Mato and Cesar Capillas and Eli Kroumova and Svetoslav Ivantchev and Gotzon Madariaga and Asen Kirov and Hans Wondratschek},
  pages       = {15--27},
  volume      = {221},
  number      = {1},
  journal     = {Zeitschrift für Kristallographie - Crystalline Materials},
  doi         = {doi:10.1524/zkri.2006.221.1.15},
  year        = {2006}
}

@article{Aroyo2006-bi2,
  author   = {Aroyo, Mois I. and Kirov, Asen and Capillas, Cesar and Perez-Mato, J. M. and Wondratschek, Hans},
  title    = {{Bilbao Crystallographic Server. II. Representations of crystallographic point groups and space groups}},
  journal  = {Acta Cryst. A},
  year     = {2006},
  volume   = {62},
  number   = {2},
  pages    = {115--128},
  month    = {Mar},
  doi      = {10.1107/S0108767305040286}
}

@article{elcoro2021magnetic,
  title={Magnetic topological quantum chemistry},
  author={Elcoro, Luis and Wieder, Benjamin J and Song, Zhida and Xu, Yuanfeng and Bradlyn, Barry and Bernevig, B Andrei},
  journal={Nat. Commun.},
  volume={12},
  number={1},
  pages={5965},
  year={2021},
  doi= {10.1038/s41467-021-26241-8},
  publisher={Nature Publishing Group UK London}
}

@article{xu2020high,
  title={High-throughput calculations of magnetic topological materials},
  author={Xu, Yuanfeng and Elcoro, Luis and Song, Zhi-Da and Wieder, Benjamin J and Vergniory, Maia G and Regnault, Nicolas and Chen, Yulin and Felser, Claudia and Bernevig, B Andrei},
  journal={Nature},
  volume={586},
  number={7831},
  pages={702--707},
  year={2020},
  doi={10.1038/s41586-020-2837-0},
  publisher={Nature Publishing Group UK London}
}

@article{yonezawa2018nematic,
  title={Nematic {Superconductivity} in {Doped} {Bi}\textsubscript{2}{Se}\textsubscript{3} {Topological} {Superconductors}},
  author={Yonezawa, Shingo},
  journal={Condens. Matter},
  volume={4},
  number={1},
  pages={2},
  year={2018},
  publisher={MDPI},
  doi={10.3390/condmat4010002}
}

@article{sasaki2012odd,
  title = {Odd-{Parity} {Pairing} and {Topological} {Superconductivity} in a {Strongly} {Spin-Orbit} {Coupled} {Semiconductor}},
  author = {Sasaki, Satoshi and Ren, Zhi and Taskin, A. A. and Segawa, Kouji and Fu, Liang and Ando, Yoichi},
  journal = {Phys. Rev. Lett.},
  volume = {109},
  issue = {21},
  pages = {217004},
  numpages = {5},
  year = {2012},
  month = {Nov},
  publisher = {American Physical Society},
  doi = {10.1103/PhysRevLett.109.217004}
}

@article{wang2016observation,
  title={Observation of superconductivity induced by a point contact on 3{D} {Dirac} semimetal {Cd}\textsubscript{3}{As}\textsubscript{2} crystals},
  author={Wang, He and Wang, Huichao and Liu, Haiwen and Lu, Hong and Yang, Wuhao and Jia, Shuang and Liu, Xiong-Jun and Xie, XC and Wei, Jian and Wang, Jian},
  journal={Nature Mater.},
  volume={15},
  number={1},
  pages={38--42},
  year={2016},
  publisher={Nature Publishing Group UK London},
  doi={10.1038/nmat4456}
}

@article{aggarwal2016unconventional,
  title={Unconventional superconductivity at mesoscopic point contacts on the 3{D} {Dirac} semimetal {Cd}\textsubscript{3}{As}\textsubscript{2}},
  author={Aggarwal, Leena and Gaurav, Abhishek and Thakur, Gohil S and Haque, Zeba and Ganguli, Ashok K and Sheet, Goutam},
  journal={Nature Mater.},
  volume={15},
  number={1},
  pages={32--37},
  year={2016},
  publisher={Nature Publishing Group UK London},
  doi={10.1038/nmat4455}
}

@article{wang2017discovery,
  title={Discovery of tip induced unconventional superconductivity on {Weyl} semimetal},
  author={Wang, He and Wang, Huichao and Chen, Yuqin and Luo, Jiawei and Yuan, Zhujun and Liu, Jun and Wang, Yong and Jia, Shuang and Liu, Xiong-Jun and Wei, Jian and Wang, Jian},
  journal={Science Bulletin},
  volume={62},
  number={6},
  pages={425--430},
  year={2017},
  publisher={Elsevier},
  doi={10.1016/j.scib.2017.02.009}
}

@article{he2013full,
  title = {Full superconducting gap in the doped topological crystalline insulator {Sn}${}_{0.6}${In}${}_{0.4}${Te}},
  author = {He, L. P. and Zhang, Z. and Pan, J. and Hong, X. C. and Zhou, S. Y. and Li, S. Y.},
  journal = {Phys. Rev. B},
  volume = {88},
  issue = {1},
  pages = {014523},
  numpages = {4},
  year = {2013},
  month = {Jul},
  publisher = {American Physical Society},
  doi = {10.1103/PhysRevB.88.014523}
}

@article{geier2020symmetry,
  title = {Symmetry-based indicators for topological {Bogoliubov}--de {Gennes} {Hamiltonians}},
  author = {Geier, Max and Brouwer, Piet W. and Trifunovic, Luka},
  journal = {Phys. Rev. B},
  volume = {101},
  issue = {24},
  pages = {245128},
  numpages = {41},
  year = {2020},
  month = {Jun},
  publisher = {American Physical Society},
  doi = {10.1103/PhysRevB.101.245128}
}

@article{kobayashi2015topological,
  title = {Topological {Superconductivity} in {Dirac} {Semimetals}},
  author = {Kobayashi, Shingo and Sato, Masatoshi},
  journal = {Phys. Rev. Lett.},
  volume = {115},
  issue = {18},
  pages = {187001},
  numpages = {5},
  year = {2015},
  month = {Oct},
  publisher = {American Physical Society},
  doi = {10.1103/PhysRevLett.115.187001}
}

\end{document}